\documentclass[prl,twocolumn,showpacs,superscriptaddress,amsmath,amssymb]{revtex4}

\usepackage{graphicx}%
\usepackage{dcolumn}%

\begin{document}

\title{Parity of the Pairing Bosons in a High-Temperature Superconductor.}

\author{S. V. Borisenko}
\affiliation{Leibniz-Institute for Solid State Research, IFW-Dresden, P.O.Box 270116, D-01171 Dresden, Germany}

\author{A. A. Kordyuk}
\affiliation{Leibniz-Institute for Solid State Research, IFW-Dresden, P.O.Box 270116, D-01171 Dresden, Germany}
\affiliation{Institute of Metal Physics of National Academy of Sciences of Ukraine, 03142 Kyiv, Ukraine}

\author{A. Koitzsch}
\author{J. Fink}
\author{J. Geck}
\author{V. Zabolotnyy}
\author{M. Knupfer}
\author{B. B\"uchner}

\affiliation{Leibniz-Institute for Solid State Research, IFW-Dresden, P.O.Box 270116, D-01171 Dresden, Germany}

\author{H. Berger}
\affiliation{Institute of Physics of Complex Matter, EPFL, CH-1015 Lausanne, Switzerland}

\author{M. Falub}

\affiliation{Swiss Light Source, Paul Scherrer Institut, CH-5234 Villigen, Switzerland}
\affiliation{Ecole Polytechnique Federale de Lausanne (EPFL), Laboratoire de Spectroscopie Electronique, Ch-1015, Lausanne, Switzerland}

\author{M. Shi}
\author{J. Krempasky}
\author{L. Patthey}
\affiliation{Swiss Light Source, Paul Scherrer Institut, CH-5234 Villigen, Switzerland}

\date{10 November 2004}
\begin{abstract}
We report the observation of a novel effect in the bilayer Pb-Bi2212 high-T$_{c}$ superconductor by means of angle-resolved photoemission with circularly polarized excitation. Different scattering rates, determined as a function of energy separately for the bonding and antibonding copper-oxygen bands, strongly imply that the dominating scattering channel is {\it odd} with respect to layer exchange within a bilayer. This is inconsistent with a phonon-mediated scattering and favours the participation of the odd collective spin excitations in the scattering mechanism in near-nodal regions of the {\bf k}-space, suggesting a magnetic nature of the pairing mediator.
\end{abstract}

\maketitle
The search for the mechanism of high-T$_{c}$ superconductivity has converged to a choice between the electron-lattice and electron-electron interactions bringing to the forefront phonons and collective spin excitations as the most probable candidates responsible for pairing. Anomalies in the electronic spectral function due to the coupling to bosonic modes are the central subject of recent ARPES studies on different kinds of high-temperature superconductors (HTSC) \cite{Johnson,Gromko,Boris_prl03,Cuk,Lanzara,Zhou}. It turned out to be possible to extract characteristics of the mode considering the peculiarities of the momentum, doping and temperature dependence of the ARPES lineshape. In such a way it was shown that the electrons couple to a mode with an energy of ~40 meV \cite{Norman,Campuz} pointing either to the spin-1 resonance observed in the neutron scattering \cite{Rossat,Mook,Fong_prl95} or to a B1g phonon mode \cite{Cuk}. Strong anisotropy of the coupling with a "hot" region near ($\pi$, 0) is interpreted either as due to the magnetic resonance \cite{Gromko,Norman,Campuz} or a combination of the buckling and breathing phonon modes \cite{Devereaux}. A challenge for the phonon scenario remains the very pronounced doping dependence of the coupling near the ($\pi$, 0) \cite{Boris_prl03,Kim}, which is easily reconciled with the spin excitations. Moreover, a very recent analysis of the nodal spectra \cite{Kord_prl04} shows, in agreement with previous data \cite{Johnson}, that the temperature and doping dependence can be well understood within the framework of a scattering mediated by the spin excitations. Again, there is an alternative interpretation of the data based on the coupling to phonons \cite{Lanzara,Zhou}. Thus, the experimental information, which allows distinguishing between these two mechanisms, is of vital importance for the field today.

In this Letter we determine the parity of the dominant scattering using ARPES with circularly polarized excitation. We consider an intermediate region in the Brillouin zone, close to the nodal direction, but still sufficiently away from it to be able to clearly resolve the bilayer splitting. We find that the scattering in this region is mediated by odd, with respect to the layer indices, excitations over a large portion of the phase diagram of a double-layered HTSC. Such a strong interband scattering is naturally understood in terms of coupling to the odd collective spin excitations.

Photoemission experiments have been performed at the Surface and Interface: Spectroscopy (SIS) beamline \cite{Flechsig} of the Swiss Light Source using a Scienta 2002 spectrometer \cite{sls}. We used circularly polarised (CP) light having the photon energy 50.4 eV, and by running the two twin electromagnet undulators in phase-matched mode \cite{Schmidt}. The typical energy and momentum resolution were of about 10 meV and 0.2$^{\circ}$. Samples were positioned by a cryo-manipulator and measured at a temperature of 20K (unless specified otherwise) - deep in the superconducting state. The storage ring has been operated in the "top-up" mode with fast orbit feedback systems \cite{Boege}, which implies the constant photon flux at the sample site and highly stable conditions between the different polarization measurements.

Intensity distribution measured in ARPES as a function of energy and momentum provides a direct access to the spectral function. We measure here such distribution twice using the circularly polarised light of positive and negative helicities. It was shown before \cite{Boris_prb04} that the emission from the bonding and antibonding bands is sensitive to the helicity of the incoming radiation. We use this property here to distinguish between closely separated features near the nodal direction. In Fig. 1a) and b) we present the spectra taken along the cut making 11 degrees with the nodal direction, as indicated in the inset of panel c), using the CP light of both helicities. Shown in the same color scale, the raw (unnormalized) data depict the distributions of the photoelectrons landing on the detector in two successive runs ($\sim$15 min each). The only experimental parameter that varies is the helicity of the incoming radiation. Already visual inspection of both plots reveals clear distinctions in ({\bf k}, $\omega$)-distributions of the photocurrent. Apart from the different structure near the Fermi level one notices the longer "tail" on the panel a) and higher intensity of the maximum in panel b).
\begin{figure}[t!]
\includegraphics[width=8.47cm]{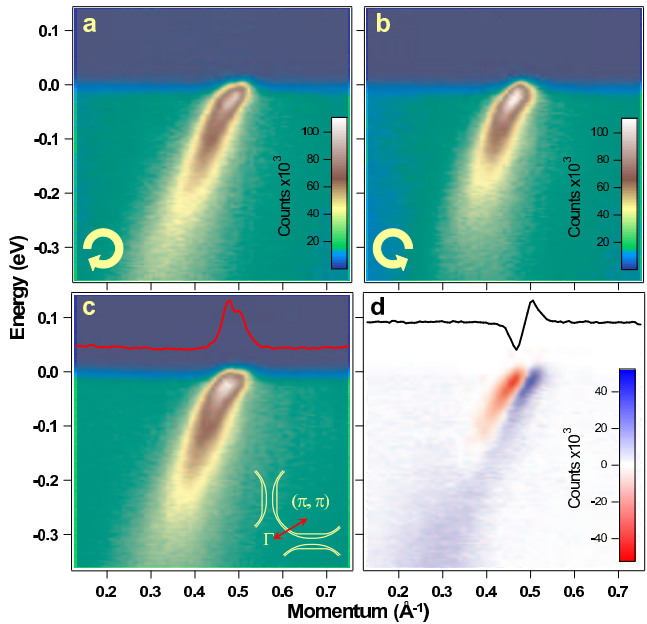}%
\caption{\label{First} ARPES spectra taken using the circularly polarized light of a) positive and b) negative helicity. c) Their sum (color plot) and $E_{F}$ - momentum distribution curve (red line). d) Their difference (color plot) and $E_{F}$ - momentum distribution (black line). The inset in panel c) shows the location in the {\bf k}-space.} 
\end{figure}
Panels c) and d) of Fig. 1 help to realize how much different the spectra actually are and how to understand this difference qualitatively. In panel c) we show the sum of the two intensity maps which can be thought of as a map taken using unpolarised light and which clearly signals the presence of the two features  - bilayer split bonding and antibonding components of the electronic structure of the double-layered Bi2212. The momentum distribution curve (MDC) at the Fermi level has a typical, well-resolved double-peak structure. The character of the dichroic signal (Fig. 1d) triggers the explanation proposed in our earlier study \cite{Boris_prb04} - depending on whether the right- or left-hand CP light impinges upon the surface, the bonding or antibonding states are selectively excited and thus dominate in the intensity distribution. However, we draw attention to the pronounced asymmetry (predominance of blue color in Fig. 1d) of the dichroic signal at higher ($>$100meV) binding energies - the consequence of the longer "tail" in the spectra from panel a). Exactly this asymmetry is the characteristic feature of the effect we are discussing in the remainder of this Letter.

At 50.4 eV excitation photon energy used here, the different helicities mainly highlight one of the bilayer split bands while the contribution from the other is minor. In order to completely eliminate the influence of the bilayer splitting, we subtract this minor contribution as follows. From Fig. 1a it is seen that the stronger component is the bonding band which crosses the Fermi level at larger momentum: we use the convention, that the bonding Fermi surface is the smaller one, i.e. the one, which is closer to the ($\pi$, $\pi$)-point (see inset in Fig. 1c). We therefore subtract from it the intensity from the Fig. 1b (where the antibonding band dominates), weighted by a factor ($\alpha$), which should account for the degree of the admixture of the weaker band and presumably depend on the degree of the circular polarisation, experimental geometry and photon energy. To determine this factor we minimize the residuals when fitting all MDC's of the resulting difference with Lorentzians, which are expected to represent the spectral function of a single feature as previous studies suggest \cite{Kord_prb05}. It turned out that there is a well-defined factor ($\alpha$=0.282, see inset to Fig. 2a) at which the errors are minimal, i.e. the lineshapes are closest to Lorentzians. The result is shown in Fig. 2a and thus represents the photocurrent corresponding purely to the bonding states. The same criterion has been applied to eliminate the admixture of the bonding band to the spectra taken with left-hand CP light (Fig. 2b). Remarkably, the determined factor ($\alpha$=0.258, see inset to Fig. 2b) is very close to the previous one indicating the high selectivity of the excitation process and approximate equality of the photoemission from both bands upon hypothetical use of unpolarised light.
\begin{figure}[t!]
\includegraphics[width=8.7 cm]{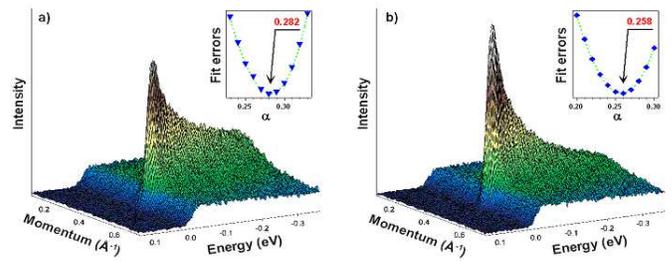}%
\caption{\label{Second} Spectra from Fig. 1 after mutual subtraction with the corresponding weighting factor (shown in the insets) which correspond to the a) bonding and b) antibonding bands.}
\end{figure}

Now, when the bilayer splitting related effects are minimized and we can directly compare the intrinsic lineshapes, we turn to the main subject of this study. One can easily show that the MDC's height when plotted as a function of energy is inversely proportional to the imaginary part of the electron self-energy. We note that unlike commonly used for this purpose MDC's widths, the height is more directly related to the imaginary part since no additional assumptions as for the bare dispersion is needed (although the determination of the absolute values is more complicated). Intensity plots shown in Fig. 2 reveal a clear-cut difference in the scattering processes of the bonding and antibonding photoholes. Near the Fermi level the scattering rate of the bonding states is higher than for the antibonding states as the corresponding MDC's are lower. At binding energies 150-350 meV the situation is inverse. Reciprocal heights of the MDC's are plotted in Fig. 3a) as dashed lines. Near the Fermi level one can restore the true behaviour dividing out the influence of the Fermi function. We show also the MDC's widths in Fig. 3b). The difference is not that pronounced but the qualitative picture is the same. There is a crossover region between 50 and 100 meV, outside of which the two curves start to diverge. Panel c) of Fig. 3, where four raw MDC's are shown, brings additional evidence for the different scattering rates near the Fermi level. Indeed, at higher binding energies both MDC's are only slightly asymmetric, and so is the MDC measured with negative helicity close to the chemical potential. The other one, however, clearly shows a double-peak structure, indicating that the antibonding component became stronger and sharper at low energies and is clearly seen even in the spectrum taken with right-hand CP light, favourable for the bonding band. This is expected if the scattering rate is noticeably lower for the one band than for the other. 
\begin{figure}[t!]
\includegraphics[width=8.47cm]{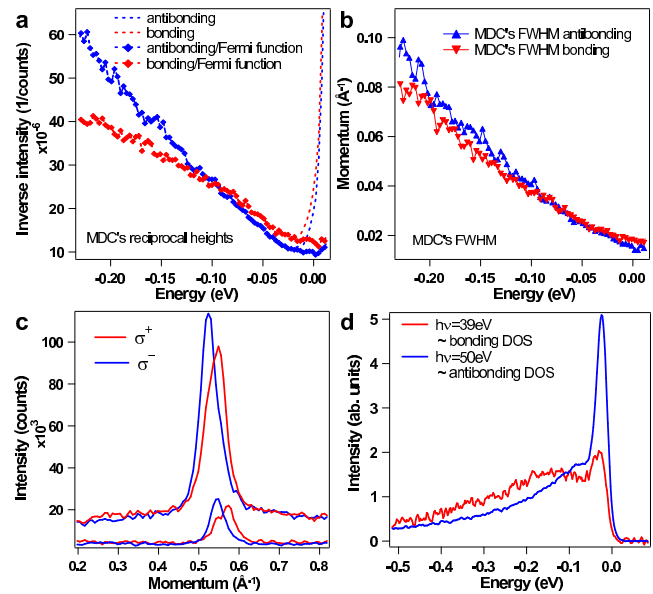}%
\caption{\label{Third} a) MDC heights as functions of energy. b)MDC widths as functions of energy. c) Two pairs of MDC's from Fig. 1a,b taken at -31 meV (upper pair) and 8 meV (lower pair) showing changing difference in the scattering rates. d) ($\pi$, 0)-spectra taken at two photon energies at which one of the bands is suppressed. Background is subtracted and spectra are normalized to the integrated intensity.} 
\end{figure}

What could be the reason for such a difference in the scattering rates of the bonding and antibonding states? From the general point of view, it is difficult to expect a fundamentally different scattering mechanism for the split pair of bands of the same atomic character. In this region of the {\bf k}-space, the difference between bare bands is negligible as tight-binding based consideration demonstrates. In the superconducting state both bands support similarly gapped Fermi surfaces \cite{Boris_prb02}. On the other hand, the scattering rate of the photohole is also proportional to the electronic density of states (DOS). Presented data suggest that the densities of those electrons which preferably fill in the hole in the bonding or antibonding bands are different. In Fig. 3d we plot the curves which represent the bonding (B) and antibonding (A) densities of states in the superconducting state. It is known that the two components of the electronic structure of bilayer Bi2212 are similar in the most of the Brillouin zone, being noticeably different only near the ($\pi$, 0)-point. Therefore we have selected two characteristic ($\pi$, 0)-spectra, which due to the careful choice of the excitation energy \cite{Boris_prl03} correspond purely to the A and B bands, as representatives of the A and B DOS. Remarkably, the curves cross at approximately the same energy ~100meV where the scattering rates are comparable (see Fig. 3a,b). This crossing is indeed expected: the binding energies of the A and B saddle points, which contribute dominantly to the DOS, are known to be at 10 - 35 meV and 150 - 200 meV respectively \cite{Boris_prl03,Kord_prl02}. Note that at low binding energy, where we found a stronger scattering for the bonding band, the DOS of the antibonding band is higher and at higher binding energies the situation is reversed.

These results now bring us to the conclusion that the scattering of electrons in the superconducting state has a well-pronounced odd character. This means that the interband (between B and A) scattering is significantly stronger compared to the intraband (within B or A) scattering. If the reversed were true, the curves in Fig. 3a,b should have been swapped over. In accord with our previous studies we consider two channels of the scattering: bosonic and fermionic \cite{Kord_prl04}. Simple symmetry-based consideration may be applied to demonstrate that in the fermionic channel the probabilities of the interband- and intraband scattering events within the framework of the direct electron-electron interaction (Auger decay) are equivalent. This consideration includes the analysis of the parity of the initial and final states with respect to the layers exchange. It turns out that even if the parity conservation is taken into account, the mentioned probabilities are both defined by the total (A+B) density of states. Thus, it is the bosonic channel, which is responsible for the observed effect. The odd scattering implies that the bosonic excitations, which mediate the scattering, must also have an odd parity with respect to the exchange between adjacent copper oxide layers. One may consider the possibility that a phonon mode can mediate the odd scattering. The so-called "half-breathing" mode has been suggested to be responsible for the dispersion anomalies near the nodal region \cite{Devereaux}. Taking into account its essentialy in-plane character one cannot reconcile the obtained results with the considerable coupling of the electrons to such a mode. There is no indication that such phonon-mediated scattering will demonstrate any imbalance with respect to the parity. On the other hand, it is known that the spectrum of collective spin excitations is dominant in the odd channel \cite{Rossat,Mook,Fong_prl95,Fong,Pailhes} thus possesing exactly the required property. Its resonance part is detected also in the overdoped samples and above T$_{c}$ in the pseudogap regime always prevailing in the odd channel in comparison with the even one. Moreover, a similar scenario of odd scattering has been invoked to successfully account for the dispersion anomalies extracted from ARPES experiments near the ($\pi$, 0)-point \cite{Eschrig}, where the famous spin-1 resonance has played the role of the scattering boson. 

To test the relevance of the observed effect to the spectrum of the magnetic excitations we have further varied three parameters: location in the {\bf k}-space, temperature and concentration of the charge carriers in the sample. The results are briefly summarized in Fig. 4. Persistance of the effect, although weakened, in the pseudogap state as well as in the superconducting state of the overdoped sample is obvious, thus tracking the characteristic behaviour of the magnetic excitations spectrum. Although little is known as for the parity of the spin excitation spectrum in the normal state from the experiments \cite{Keimer}, the theoretical estimates clearly point to its odd character \cite{Dahm}. Moreover, according to the expectations, the effect must be even less pronounced in the normal state of the overdoped samples. Our data taken with different experimental setup suggest that this is indeed the case \cite{Boris_prb04}. Momentum-dependent data (Fig. 4c) conclude this detailed inventory of the properties demonstrating another magnetism-related feature - the difference becomes more pronounced when moving towards the antinodal region. We note that upon approaching the nodal direction, the dominance of the bonding band at higher binding energies does not disappear (seen as blue or red colour in lower parts of the dichroism patterns), however, the dichroism itself gets weaker as it is expected to vanish in the mirror plane \cite{Boris_prl04,Boris_nature}.

\begin{figure}[t!]
\includegraphics[width=8.47cm]{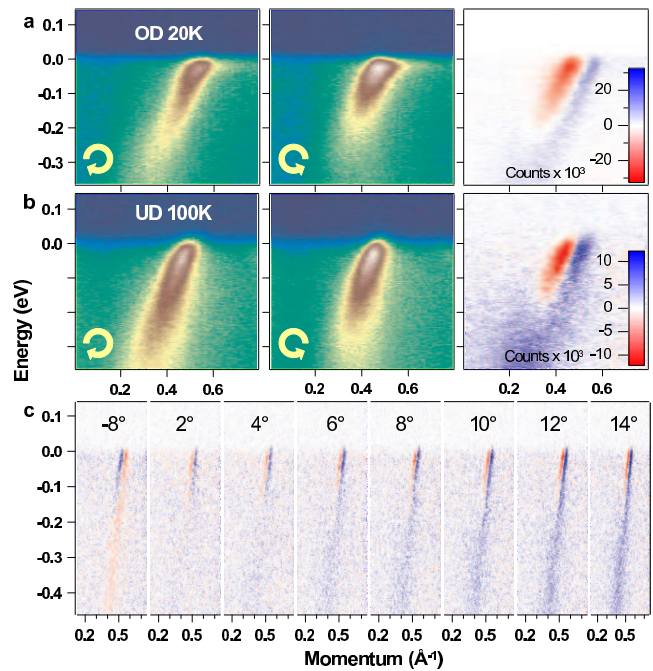}%
\caption{\label{Fourth} a) Presence of the effect in an overdoped (T$_{c}$=72K) compound. b) Presence of the effect in the pseudogap regime of an underdoped sample (T$_{c}$=76K). c) Momentum dependence of the effect.} 
\end{figure}

The strength of the coupling between electrons and the antiferromagnetic spin fluctuations drastically increases for those electrons which are separated in the reciprocal space by the vector ($\pi$, $\pi$). We point out that this condition is, in principle, fulfilled for all points on the Fermi surface if one recalls about the "shadow" FS recently shown to be an intrinsic feature of the photoemission from the copper oxide planes \cite{Koitzsch}. Its presence implies that the vector ($\pi$, $\pi$) becomes a reciprocal lattice vector, which is not surprising since a certain degree of buckling is characteristic to the CuO plane. Thus, some of the high-T$_{c}$ cuprates emerge as systems, which are "structurally prepared" for the strong interplay between electrons and spin fluctuations.

We attribute the observed effect to the interband scattering mediated by the collective spin excitations. Inelastic neutron scattering experiments have demonstrated that such spin excitations possess all essential properties of the pairing mediating excitations - they are universally present in copper oxides, have a suitable energetics and present at T$_{c}$. Our results indicate that exactly interaction with the spin excitations captures all characteristic features of the low energy electron dynamics in bilayer cuprates. Magnetic excitations therefore fulfill the last and crucial requirement - strong coupling to the conduction electrons, and thus emerge as the most probable candidate for mediation the electron pairing in superconducting cuprates.

The project is part of the Forschergruppe FOR538 and was supported by the DFG under Grant No. KN393/4. This work was performed at the Swiss Light Source, Paul Scherrer Institut, Villigen, Switzerland. We thank R. H\"ubel for technical support.

\end{document}